\documentclass[12pt]{article}
\usepackage{graphicx}

\begin{document}

\title{The structural de-correlation time:  A robust statistical measure of convergence of biomolecular simulations}
\author{Edward Lyman\footnote{elyman@ccbb.pitt.edu},
and Daniel M. Zuckerman\footnote{dmz@ccbb.pitt.edu}\\
Department of Computational Biology, School of Medicine,\\ 
3088 BST3, 3501 Fifth Ave., University
of Pittsburgh, Pittsburgh, PA 15213}
\date{\today}
\maketitle
\begin{abstract}
Although atomistic simulations of proteins and other biological systems are approaching 
microsecond timescales, the quality of trajectories has remained difficult to assess.  
Such assessment is critical not only for establishing the relevance of any individual 
simulation but also in the extremely active field of developing computational methods.  
Here we map the trajectory assessment problem onto a simple statistical calculation of 
the ``effective sample size'' - i.e., the number of statistically independent configurations.  
The mapping is achieved by asking the question, ``How much time must elapse between snapshots 
included in a sample for that sample to exhibit the statistical properties expected for 
independent and identically distributed configurations?''  The resulting ``structural 
de-correlation time'' is robustly calculated using exact properties deduced from our previously 
developed ``structural histograms,'' without any fitting parameters. 
We show the method is equally and directly applicable 
to toy models, peptides, and a 72-residue protein model.  Variants of our approach can readily 
be applied to a wide range of physical and chemical systems.
\end{abstract}

What does convergence mean?  The answer is not simply of abstract interest, since many aspects 
of the biomolecular simulation field depend on it.  
When parameterizing 
potential functions, it is essential to know whether inaccuracies are attributable to the 
potential, rather than under-sampling. 
In the extremely active area 
of methods development for equilibrium sampling, it is necessary to demonstrate that a novel 
approach is better than its predecessors, in the sense that it equilibrates 
the relative populations of different conformers in less CPU time\cite{repex-note}. 
And in the important 
area of free energy calculations, under-sampling can result in both systematic error 
and poor precision.      

To rephrase the basic question, given a simulation trajectory (an ordered set of correlated 
configurations), what characteristics should be observed if convergence has been achieved?  
The obvious, if tautological, answer is that all states should have been visited with the 
correct relative probabilities, as governed by a Boltzmann factor (implicitly, free energy) 
in most cases of physical interest.  Yet given the omnipresence of statistical error, it has 
long been accepted that such idealizations are of limited value. The more pertinent questions 
have therefore been taken to be: Does the trajectory give reliable estimates for quantities 
of interest?  What is the statistical uncertainty in these 
estimates\cite{binder-book,ferrenberg,flyvbjerg,frenkel-book}? In other words, \emph{convergence 
is relative}, and in principle, it is rarely meaningful to describe a simulation as not converged, 
in an absolute sense.  
(An exception is when a priori information indicates the trajectory has failed to visit certain states.)

Accepting the relativity of convergence points directly to the importance of computing statistical 
uncertainty. The reliability of ensemble averages typically has been gauged in the context of basic 
statistical theory, by noting that statistical errors decrease with the square-root of the number of 
independent samples. 
The number of independent samples $\cal{N} _{A}$ pertinent to the uncertainty in a quantity $A$, in turn, has been 
judged by comparing the trajectory length to the timescale of $A$'s correlation with itself---$A$'s autocorrelation 
time\cite{binder-optimum-accept,ferrenberg,frenkel-book}.  Thus, a trajectory of 
length $t_{sim}$ with an 
auto-correlation time for $A$ of $\tau_A$ can be said to provide an estimate for $A$ with relative precision of 
roughly $\sqrt{(1/\cal{N} _A)} \sim \sqrt{( 2\tau_A / t_{sim} )}$.

However, the estimation of correlation times can be an uncertain business, as good measurements 
of correlation functions require a lot of data\cite{zwanzig-finite-error,go-finite-error}.  
Furthermore, different quantities typically have different correlation times. 
Other assessment approaches have therefore been proposed, such as the 
ergodic measure\cite{straub-pnas-93,thirumalai-jacs93}, analysis of principle components\cite{hess-pre02} 
and, more recently, structural 
histograms\cite{el-dmz-class05}.   Although these approaches are applicable to quite complex systems 
without invoking correlation functions, they attempt only to give an overall sense of convergence 
rather than quantifying precision.

Flyvbjerg and  Petersen provided perhaps the most satisfying approach to quantifying the precision 
of an estimate in any particular quantity without relying on correlation functions\cite{flyvbjerg}.  
The sophisticated block averaging scheme they present gauges whether correlation effects have been removed by 
considering a range of block sizes.  The reasoning underlying their analysis is that, once the block 
length is longer than any correlation time(s), the estimated precision (statistical uncertainty) will 
no longer depend on block size.

Our approach generalizes the logic implicit in the Flyvbjerg-Petersen analysis by developing an overall 
structural de-correlation time which can be estimated, simply and robustly, in biomolecular and other 
systems.   The key to our method is to view a simulation as sampling an underlying distribution 
(typically a Boltzmann factor) of the configuration space, from which all equilibrium quantities follow.  
Our approach builds implicitly on the multi-basin picture proposed by Frauenfelder and 
coworkers\cite{frau-biochem75,frau-science91}, in which conformational equilibration requires 
equilibrating the relative populations of the various conformational substates.

On the basis of the configuration-space distribution, we can define the general effective sample size 
$\cal{N}$ and the associated (de-)correlation time $\tau_{\rm{dec}}$ associated with a particular trajectory.  
Specifically, $\tau_{\rm{dec}}$ is the minimum time that must elapse between configurations for them to become 
fully decorrelated (i.e., with respect to any quantity).  Here, fully decorrelated has a very specific 
meaning, which leads to testable hypotheses: a set of fully decorrelated configurations will exhibit 
the statistics of an independently and identically distributed (i.i.d.) sample of the governing 
Boltzmann-factor distribution.  Below, we detail the tests we use to compute $\tau_{\rm{dec}}$, which build on our 
recently proposed structural histogram analysis\cite{el-dmz-class05}; see also \cite{simm-hybrid-repex-jctc06}.

The key point is that the expected i.i.d.\ statistics must apply to any assay of a decorrelated sample.  
The contribution of the present paper is to recognize this, and then to describe an assay directly probing 
the configuration-space distribution for which analytic results are easily obtained  for any system,  
assuming an i.i.d.\ sample.  Procedurally, then, we simply apply our assay to increasing values hypothesized 
for $\tau_{\rm{dec}}$.  When the value is too small, the correlations lead to anomalous statistics (fluctuations), 
but once the assayed fluctuations match the analytic i.i.d. predictions, the de-correlation time $\tau_{\rm{dec}}$ has 
been reached.  Hence, \emph{there is no fitting of any kind}.  Importantly, by a suitable use of our ``structural 
histograms''\cite{el-dmz-class05}, which directly describe configuration-space distributions, we can map a system of any complexity 
to an exactly soluble model.  

In practical terms, our analysis readily computes the configurational/structural decorrelation time 
$\tau_{\rm{dec}}$ (and hence the number of independent samples $\cal{N}$) for a long trajectory many times the 
length of $\tau_{\rm{dec}}$.  In turn, this provides a means for estimating statistical uncertainties in 
observables of interest, such as relative populations.  Of equal importance, our analysis can reveal 
when a trajectory is dominated by statistical error, i.e., when the simulation time $t_{sim} \sim \tau_{\rm{dec}}$.  We note, however, 
that our analysis remains subject to the intrinsic limitation  pertinent to all methods 
which aim to judge the quality of conformational sampling---of not knowing 
about parts of configuration space never visited by the trajectory being analyzed.

In contrast to most existing quantitative approaches, which attempt to assess convergence of a single 
quantity, our general approach enables the generation of ensembles of known statistical properties. 
These ensembles in turn can then be used for many purposes beyond 
ensemble averaging, such as docking, or developing a better understanding of native protein ensembles. 

In the remainder of the paper, we describe the theory behind our assay, and then successfully apply it 
to a wide range of systems.  We first consider a two-state Poisson process for illustrative purposes, 
followed by molecular systems: di-leucine peptide ($2$ residues; $50$ atoms), Met-enkephalin 
($5$ residues; $75$ atoms), and a coarse-grained model of the N-terminal domain of calmodulin ($72$ united 
residues).  For all the molecules, we test that our calculation for $\tau_{\rm{dec}}$ is insensitive to details of 
the computation. 
\section{Theory}
Imagine that we are handed a ``perfect sample'' of configurations
of a protein---perfect, we are told, because it is made up of configurations that 
are fully independent of one another. How could we test this assertion? 
The key is to 
note that, for any arbitrarily defined partitioning of the sample of $N$ configurations into $S$ subsets (or bins), 
subsamples of these $N$ configurations obey very simple statistics. 
In particular, the expected
variance in the population of a bin, as estimated from many subsamples, depends only on the 
population of the bin and size $n$ of the subsample, as long as $N>>n$. 

Of course, a sample generated by a typical simulation is not made up of independent configurations. But since we know 
how the variance of subsamples should behave for an ideal sample of independent configurations, we are able to determine 
how much simulation time must elapse before configurations may be considered independent. 
We call this time the structural decorrelation time, $\tau_{\rm{dec}}$. Below, we show how to partition the 
trajectory into structurally defined subsets for this purpose, and how to extract $\tau_{\rm{dec}}$.

There is some precedence for using the  populations of structurally defined bins as 
a measure of convergence\cite{el-dmz-class05}. Smith 
\emph{et al} considered the number of structural clusters as a function of time as 
 a way to evaluate the breadth and convergence of conformational sampling, and found this to be a much 
more sensitive indicator of sampling than other commonly used measures\cite{vangun-cluster02}. 
Simmerling and coworkers went one step further, and compared the populations of the clusters as sampled by 
different simulations\cite{simm-hybrid-repex-jctc06}. Here, we go another step, by noting 
that the statistics of populations of structurally defined bins provide a unique 
insight into the quality of the sample.

Our analysis of a simulation trajectory proceeds in two steps, both described in 
Sec.~\ref{methods}:

\parbox{5in}{
(I) A structural histogram is constructed. The 
histogram is a unique classification (a binning, not a clustering) of the trajectory based upon a set of 
reference structures, which are selected at random from the trajectory. 
The histogram so constructed defines a discrete probability distribution, $P(S)$, indexed by the 
set of reference structures $S$.\\ 
(II) We consider different subsamples of the trajectory, defined by a fixed interval of simulation time 
$t$. A particular ``$t$ subsample'' of size $n$ is formed by pulling $n$ frames in sequence  
separated by a time $t$ (see Fig.~\ref{subsamplefig}). When $t$ gets large enough, it is as if we 
are sampling randomly from $P(S)$. The smallest such $t$ we identify as the structural decorrelation time, 
$\tau_{\rm{dec}}$, as explained below. 
}

\subsection{Structural Histogram\label{sec-histo}}
A ``structural histogram'' is a one-dimensional population analysis of a trajectory based on 
a partitioning (classification) of configuration space. Such classifications are simple 
to perform based on proximity of the sampled configurations to a set of reference structures 
taken from the trajectory itself\cite{el-dmz-class05}. 
The structural histogram will 
form the basis of the decorrelation time analysis. It defines a distribution, 
which is then used to answer the question, ``How much time must elapse between 
frames before we are sampling randomly from this distribution?'' Details are given in 
Sec.~\ref{methods}.

Does the the equilibration of a structural histogram reflect the equilibration of the 
underlying conformational substates (CS)? Certainly, several CS's will be lumped 
together into the same bin, while others may be split between one or more bins. But clearly, 
equilibration of the overlying histogram bins requires equilibration of the 
underlying CS's. We will present evidence that this is indeed the case in Sec.~\ref{results}. 
Furthermore, since the configuration space distribution (and the statistical error associated 
with our computational estimate thereof) controls \emph{all} ensemble averages, it 
determines the precision with which these averages are calculated. We will 
show that the convergence of a structural histogram is very sensitive to configuration 
space sampling errors.  

\subsection{Statistical analysis of $P(S)$ and the decorrelation time $\tau_{dec}$\label{theory-stat}}
In this section we define an observable, $\sigma^2_{\rm{obs}}(t)$, which depends very 
sensitively on the equilibration of the bins of a structural histogram as a function of simulation time $t$. 
Importantly, $\sigma^2_{\rm{obs}}(t)$ can be exactly calculated for a 
histogram of fully decorrelated structures. Plotting $\sigma^2_{\rm{obs}}(t)$ as a function 
of $t$, we identify the time at which the observed value 
equals that for fully decorrelated structures as the structural decorrelation time. 

Given a trajectory of $N$ frames, we build a uniform histogram of $S$ bins $P(S)$, 
using the procedure described in Sec.~\ref{methods}. 
By construction, the likelihood that a randomly selected frame belongs to bin `$i$' of
$P$ is simply $1/S$. 
Now imagine for a moment that the trajectory was generated by an algorithm which 
produced structures that are completely independent of 
one another. Given a subsample of $n$ frames of this correlationless trajectory,  
the expected number of structures in the 
subsample belonging to a particular bin is simply $n/S$, regardless of the ``time'' separation of the
frames.

\begin{figure}
\includegraphics[scale=0.5]{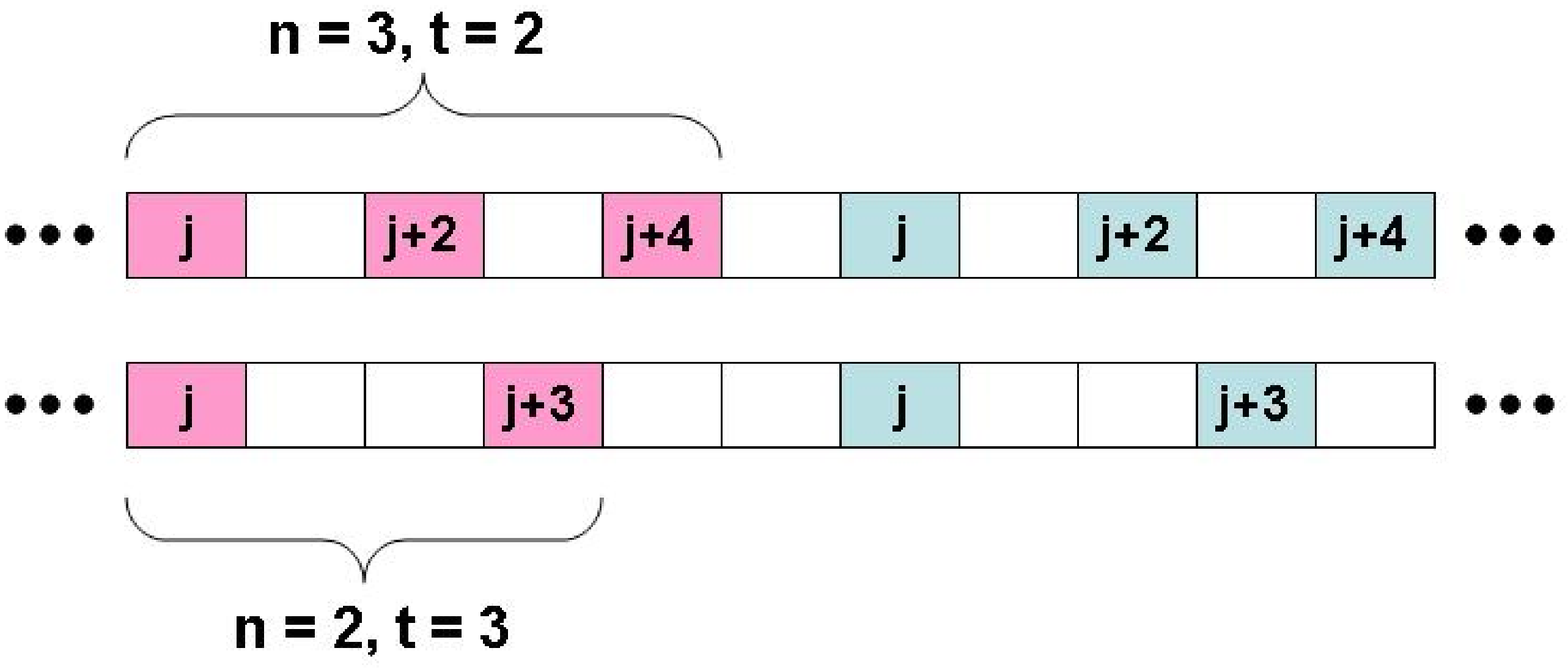}
\caption{A trajectory can be subsampled in many ways, corresponding to different subsample sizes $n$ and 
intervals $t$. In the top figure, the pink highlighted frames belong to an $n=3$, $t =2$ 
subsample, the blue frames to another subsample of the same type. 
The bottom figure shows two $n=2$, $t = 3$ subsamples. The frame index (simulation time) is labelled 
by $j$.}  
\label{subsamplefig}
\end{figure}

As the trajectory does not consist of independent structures, the statistics of subsamples 
depend on how the subsamples are selected. For example, a subsample of frames close 
together in time are more likely to belong to the same bin, as compared to a subsample 
of frames which span a longer time. Frames that are close together (in 
simulation time) are more likely to be in similiar conformational substates, while 
frames separated by 
a time which is long compared to the typical inter-state transition times 
are effectively independent. The difference between these two 
types of subsamples---highly correlated vs. fully independent---is reflected in the 
\emph{variance} among a set of subsampled bin populations (see Fig.~\ref{subsamplefig}). Denoting the population of bin 
$i$ obseved in subsample $k$ as $m_i^k$, the fractional population $f_i^k$ is 
defined as $f_i^k\equiv m_i^k /n$. The variance $\sigma^2(f_i)$in the fractional population $f_i$ of bin $i$ is then 
defined as
\begin{equation} 
\sigma^2(f_i)\equiv \overline{(f^k_i-\overline{f_i})^2}, 
\label{variancedef}
\end{equation}
where overbars denote averaging over subsamples: $\overline{f_i} = \frac{1}{N}\sum_{k=1}^N f_i^k$. Since 
here we are considering only uniform probability histograms, $\overline{f_i}$ is the same 
for every $i$: $\overline{f_i}=f=1/S$.

The expected variance of bin populations when allocating $N$ \emph{fully independent} structures 
to $S$ bins is calculated in introductory 
probability texts under the rubric of ``sampling without replacement\cite{prob-book}.'' 
The variance in fractional occupancy of each bin of this (hypergeometric) distribution 
depends only on the total number of \emph{independent} structures $\cal{N}$, the size $n$ of the subsamples 
used to ``poll'' the distribution, and the fraction $f$ of the structures which are contained 
in each bin:  
\begin{equation}
\sigma^2(f) = \frac{f(1-f)}{n}\left( \frac{{\cal N}-n}{{\cal N}-1}\right) .
\label{hypervar}
\end{equation}

But can we use this exact result to infer something 
about the correlations that are present in a typical trajectory? 
Following the intuition that frames close together in time are correlated, while frames far apart are 
independent, we compute the variance in Eq.~\ref{variancedef} for different sets of subsamples, 
which are distinguished by a fixed time $t$ between subsampled frames (Fig.~\ref{subsamplefig}). We 
expect that averaging over subsamples that consist of frames close together in time 
will lead to a variance which is higher than that expected from an ideal sample (Eq.~\ref{hypervar}). 
As $t$ increases, the variance should decrease as the frames in each subsample 
become less correlated. Beyond some $t$ (provided the trajectory is long enough), the subsampled frames 
will be independent, and the computed variance will be that expected from an i.i.d.\ sample.

In practice, we turn this intuition into a (normalized) observable $\sigma^2_{\rm{obs}}(f;n,t)$ in the following way:

\parbox{5in}{
(i) Pick a subsample size $n$, typically between $2$ and $10$. Set $t$ to the time between stored configurations.\\
(ii) Compute $\sigma^2_i$ according to Eq.~\ref{variancedef} for each bin $i$.\\
(iii) Average $\sigma^2_i$ over all the bins and normalize by $\sigma^2(f)$---the 
variance of an i.i.d.\ sample (Eq.~\ref{hypervar}):\\ 
\begin{equation}
\sigma^2_{\rm{obs}}(f;n,t)=\frac{1}{S}\sum_{i=1}^S \sigma^2_i(f;n,t)/\sigma^2(f).\\
\label{sigmadef}
\end{equation}
By construction, $\sigma^2_{\rm{obs}}(f;n,t) = 1$ for samples consisting of independent frames.\\
(iv) Repeat (ii) and (iii) for increasing $t$, until the subsamples span a 
number of frames on the order of the trajectory length.\\ 
}

Plotting $\sigma^2_{\rm{obs}}(f;n,t)$ as a function of $t$, we identify the subsampling interval $t$ at which the 
variance first equals the 
theoretical prediction as the structural decorrelation time, $\tau_{\rm{dec}}$. For frames which are separated 
by at least $\tau_{\rm{dec}}$, it is as if they were drawn independently from the distribution 
defined by $P(S)$. The effective sample size $\cal{N}$ is then the number of frames $T$ in the 
trajectory divided by $\tau_{dec}$. Statistical uncertainty on thermodynamic 
averages is proportional to ${\cal N}^{-1/2}$.

But does $\tau_{\rm{dec}}$ correspond to a physically meaningful timescale? 
Below, we show that the answer to this question is affirmative, 
and that, for a given trajectory, the same $\tau_{\rm{dec}}$ is computed, regardless of the histogram. 
Indeed, $\tau_{\rm{dec}}$ does not depend on whether it is calculated based on a uniform 
or a nonuniform histogram.

\section{Results\label{results}}
In the previous section, we introduced an observable, $\sigma^2_{\rm{obs}}(f;n,t)$, and argued that it ought to be 
sensitive to the conformational convergence of a molecular simulation. However, we need to ask 
whether the results of the analysis reflect physical processes present in the simulation. After all, 
it may be that good sampling of a structural histogram is not indicative of good sampling of the 
conformation space. 

Our strategy is to first test the analysis on some models with simple, known convergence 
behavior. We then turn our attention to more complex systems, which sample multiple 
conformational substates on several different timescales.    

\subsection{Poisson process}
Perhaps the simplest nontrivial model we can imagine has two-states, with 
rare transitions between them. If we specify that the likelihood of a transition in a unit interval of 
time is a small constant $\kappa < 1$ (Poisson process), then the average lifetime of each state is simply $1/\kappa$. 
Transitions are instantaneous, so that a ``trajectory'' of this model is simply a record of which 
state (later, histogram bin) was occupied at each timestep. Our decorrelation analysis is designed to  
answer the question, ``Given that the model is in a particular state, how much time must elapse before there is an 
equal probability to be in either state?''

\begin{figure}
\includegraphics[scale=0.5]{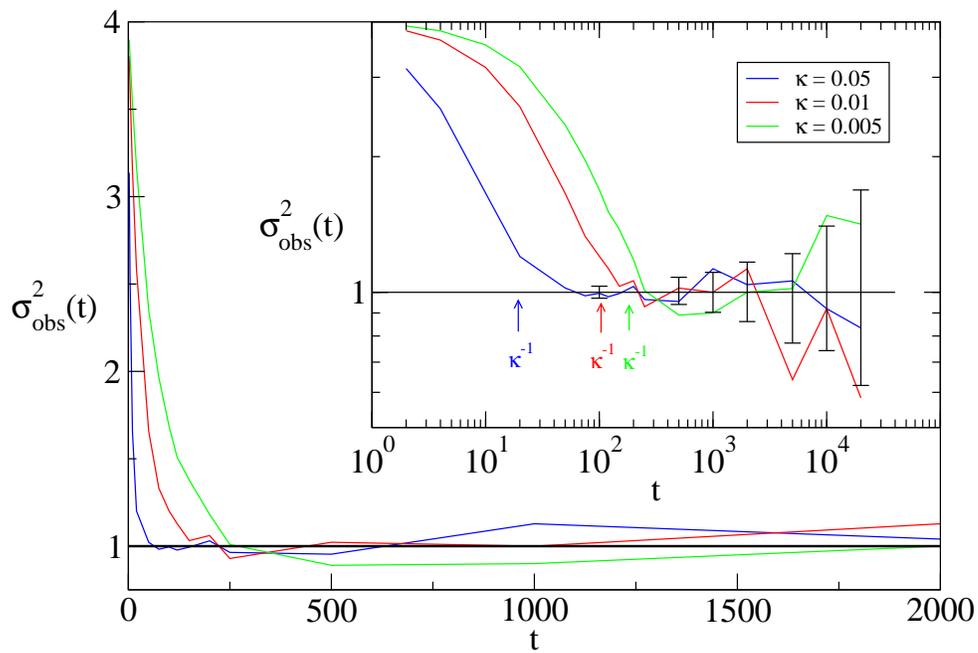}
\caption{Plotted is the behavior of $\sigma ^2_{\rm{obs}}(n,t)$ for three values of $\kappa$.
All the data have been rescaled by the variance predicted for independent sampling of the two 
states (Eq.~\ref{hypervar}).} 
\label{bernoullifig} 
\end{figure}

Figure~\ref{bernoullifig} shows the results of the analysis for several different values of $\kappa$. 
The horizontal axis measures the time between subsampled frames. Frames that are close together 
are likely to be in the same state, which results in a variance higher than that expected from 
an uncorrelated sample of the two states. As the time between subsampled frames increases, the 
variance decreases, until reaching the value predicted for independent samples, where it stays. 

The inset demonstrates that the time for which the variance first reaches the theoretical 
value is easily read off when the data are plotted on a log-log scale. In all three 
cases, this value correlates well with the (built-in) transition time $1/\kappa$. It is noteworthy 
that, in each case, we actually must wait a bit longer than $1/\kappa$ before the subsampled 
elements are uncorrelated. This likely reflects the additional waiting time necessary for the Poisson 
trajectory to have equal likelihood of being in either state.

As $t$ gets larger, the number of subsamples which ``fit'' into a trajectory decreases, and 
therefore $\sigma^2_{\rm{obs}}(t)$ is averaged over fewer subsamples. This results in some 
noise in measured value of $\sigma^2_{\rm{obs}}(t)$, which gets more pronounced with increasing $t$.  
To quantify this behavior, we
added an $80$\% confidence interval to the theoretical prediction, indicated by the error bars
in the inset of Fig.~\ref{bernoullifig}. Given an $n$ and $t$, the number of subsamples is fixed. The error
bars indicate the range where $80$ \% of variance estimates fall, based on this fixed number of
(hypothesized) independent samples from the hypergeometric distribution defined by $P(S)$. 

\subsection{Leucine dipeptide}
Our approach readily obtains the physical timescale governing conformational equilibration 
in molecular systems. 
Implicitly solvated leucine dipeptide (ACE-Leu$_2$-NME), having fifty atoms, is 
an ideal test system because
a thorough sampling of conformation space is possible by brute
force simulation. The degrees of freedom that distinguish the major 
conformations are the $\phi$ and $\psi$ dihedrals of the backbone, though 
side-chain degrees of freedom complicate the landscape by introducing 
many locally stable conformations within the major Ramachandran basins. 
It is therefore 
intermediate in complexity between a ``toy-model'' and larger peptides. 

\begin{figure}
\includegraphics[scale=0.5]{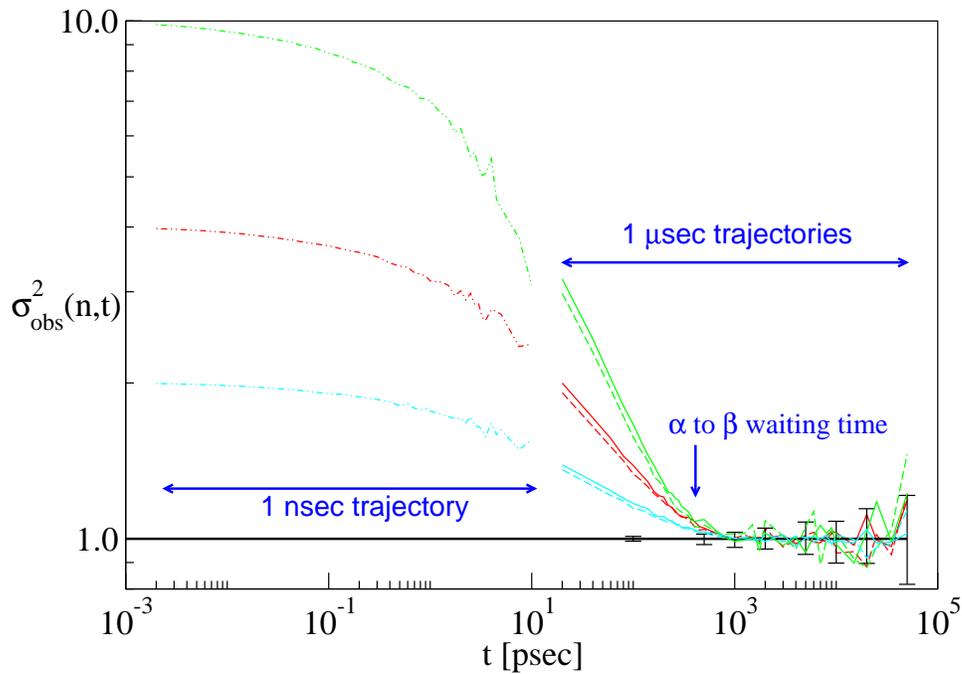}
\caption{Convergence analysis of two independent $1$ $\mu$sec dileucine trajectories (distinguished by 
dashed and solid lines) and a single $1$ nsec trajectory (dash-dot lines) for $3$ different 
subsample sizes: $n=2$ (blue), $n=4$ (red), and $n=10$ (green). 
The solid horizontal line indicates the expected variance for i.i.d.~samples. 
The average time between $\alpha\rightarrow\beta$ transitions is indicated. An $80$ \% confidence 
interval on the ($n=4$, red) theoretical prediction for independent samples is indicated by the error bars.}
\label{dileufig}
\end{figure}

Two independent trajectories of $1$ $\mu$sec each were analyzed; the simulation 
details have been reported elsewhere\cite{marty-shift-jpc05}. For each trajectory, $9$ independent 
histograms consisting of $10$ bins of uniform probability were built as described in Sec.~\ref{sec-histo}. 
For each histogram, $\sigma ^2_{\rm{obs}}(n,t)$ (Eq.~\ref{sigmadef}) was computed for $n=2,4,10$. 
We then averaged $\sigma ^2_{\rm{obs}}(n,t)$ over the $9$ independent histograms separately for 
each $n$ and each trajectory---these averaged signals are 
plotted in Fig.~\ref{dileufig}. 

When the subsamples consist of frames separated by short times $t$, 
the subsamples are made of highly correlated frames. This leads to an observed  
variance greater than that expected for a sample of independent snapshots, as calculated 
for each $n$ from Eq.~\ref{hypervar} and shown as a thick black horizontal line. $\sigma ^2_{\rm{obs}}(n,t)$ then 
decreases monotonically with time, until it matches the theoretical prediction for decorrelated snapshots 
at about $900$ psec. The agreement between the computed and theoretical variance (with no fitting parameters) 
indicates that the subsampled frames are behaving as if they were sampled at random from the structural 
histogram. We therefore identify $\tau_{\rm{dec}} = 900$ psec, giving an effective sample size of just over $1,100$ 
frames.  

Does the decorrelation time correspond to a physical timescale? First, we note that 
$\tau_{dec}$ is independent of the subsample size $n$, as shown in Fig.~\ref{dileufig}. 
Second, we note that the decorrelation 
times agree between the two independent trajectories. This is expected, since the trajectories are quite long 
for this small molecule, and therefore should be very well-sampled. Finally, the decorrelation 
time is consistent with the typical transition time between the $\alpha$ and $\beta$ basins of the Ramachandran map,
which is on the order of $400$ psec in this model. As in the Poisson process, $\tau_{\rm{dec}}$ is a bit 
longer than the $\alpha \rightarrow \beta$ transition time. 

How would the data look if we had a much shorter trajectory, of the order of $\tau_{\rm{dec}}$? This is also 
answered in Fig.~\ref{dileufig}, where we have analyzed a dileucine trajectory of only $1$ nsec 
in length. Frames every $10$ fsec, so that this trajectory had the same 
total number of frames as each of the $1$ $\mu$sec trajectories. The results 
are striking---not only does $\sigma ^2_{\rm{obs}}(n,t)$ fail to attain the value for 
independent sampling, but the values appear to connect 
smoothly (apart from some noise) with the data from the longer trajectories. (We stress that the $1$ nsec 
trajectory was generated and analyzed independently of both $1$ $\mu$sec trajectories---it is not simply the first nsec of 
either.) In the event that we had only the $1$ nsec trajectories, we could state unequivocably that they are poorly 
converged, since they fail to attain the theoretical prediction for a well-converged trajectory.

\begin{figure}
\includegraphics[scale=0.5]{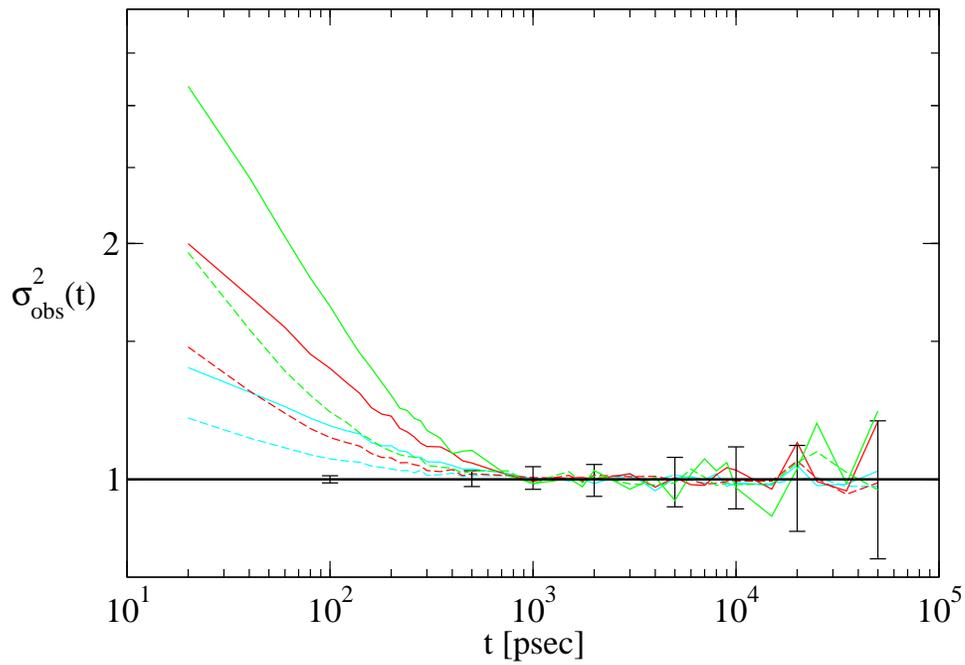}
\caption{Convergence analysis of a single $1$ $\mu$sec dileucine trajectory for different numbers 
of reference structures: $S=10$ (solid lines) and $S=50$ (dashed lines). The colors and error bars are 
the same as in the previous plot.
}
\label{diff-num-refs}
\end{figure}

We also investigated whether the decorrelation time depends on the number of reference 
structures used to build the structural histogram. As shown in Fig.~\ref{diff-num-refs}, 
$\tau_{\rm{dec}}$ is the same, whether we use a histogram of $10$ bins or $50$ bins. 
(Fig.~\ref{dileufig} used $10$ bins.) It is 
interesting that the data are somewhat smoothed by dividing up the sampled space
among more reference structures. While this seems to argue for increasing the number of 
reference structures, it should be remembered that increasing the number of references 
by a factor of $5$ increases the computational cost of the analysis by the same factor, 
while $\tau_{\rm{dec}}$ is robustly estimated based on a histogram containing $10$ bins.

\subsection{Calmodulin}
We next considered a previously developed united-residue model of the N-terminal domain of calmodulin\cite{zuckerman-calmod}. 
In the ``double native'' G{\=o} potential used, both 
the apo (Ca$^{2+}$--free)\cite{camapo-jmb92} and holo (Ca$^{2+}$--bound)\cite{bax-camholo-nature95} structures are stabilized, 
so that occasional transitions are observed between the two states. 

In contrast with the dileucine 
model just discussed, our coarse-grained calmodulin simulation has available a much larger conformation space. 
The apo-holo transition represents a motion entailing $4.6$ \AA\ RMSD, and involves a collective 
rearrangement of helices. In addition to apo-holo transitions, the trajectories include partial unfolding 
events, which do not lend themselves to an interpretation as transitions between 
well-defined states. In light of these different processes, it will be interesting to see 
how our analysis fares.
  
Two independent trajectories were analyzed, each $5.5\times 10^7$ Monte Carlo sweeps (MCS) in length. 
Each trajectory was begun in the  
apo configuration, and approximately $40$ transition events were observed in each. 
For both trajectories, the analysis was averaged over $4$ independent 
histograms, each with $10$ bins of uniform probability. 

\begin{figure}
\includegraphics[scale=0.5]{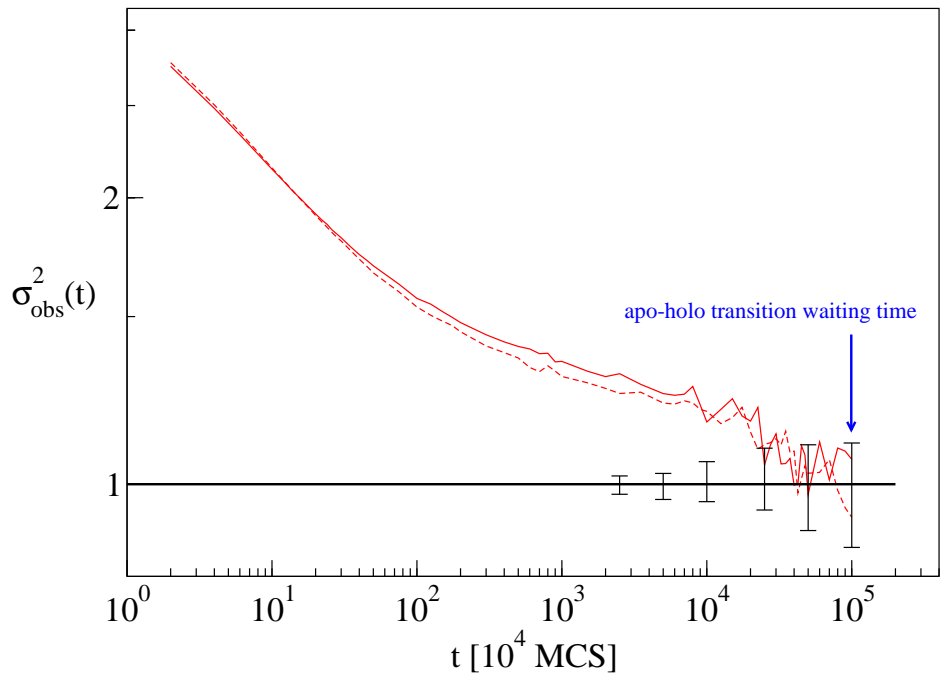}
\caption{Convergence data for two independent calmodulin trajectories. $\sigma^2(t)$ is plotted for a sample size 
of $n=4$. Error bars indicate $80$ \% confidence intervals for uncorrelated subsamples of size $n=4$.}
\label{doublegofig}
\end{figure}

The results of the analysis are shown in Fig.~\ref{doublegofig}. It is interesting that the decorrelation time 
estimated from Fig.~\ref{doublegofig} is about a factor of $2$ \emph{shorter} than the average waiting time 
between $\alpha\rightarrow\beta$ transitions. This is perhaps due to the noisier signal (as compared to the 
previous cases), which is in turn due to the small number of transition events 
observed---about $40$ in each trajectory, compared to about 
$2.5\times10^3$ events in the dileucine trajectories. Alternatively, it may be that 
there are other, longer timescale processes, such as partial unfolding and refolding events, which 
must be sampled before convergence is attained. 

In either case, our analysis yields a robust 
estimate of the decorrelation time, regardless of the underlying processes.
The conclusion we draw from this data is that one should only interepret 
the decorrelation analysis as ``logarithmically accurate'' (up to a factor of $\sim 2$) when the data are noisy. 
\subsection{Met-enkephalin}
In the previous examples, we considered models which admit a description in terms of two dominant 
states with occasional transitions between them. Here, we study the highly flexible pentapeptide met-enkephalin 
(NH$_3^+$-Tyr-[Gly]$_2$-Phe-Met-COO$^-$), which does not lend itself to such a simple description. 
Our aim is to see how our convergence analysis will perform in this case, where multiple 
conformations are interconverting on many different timescales.

\begin{figure}
\includegraphics[scale=0.5]{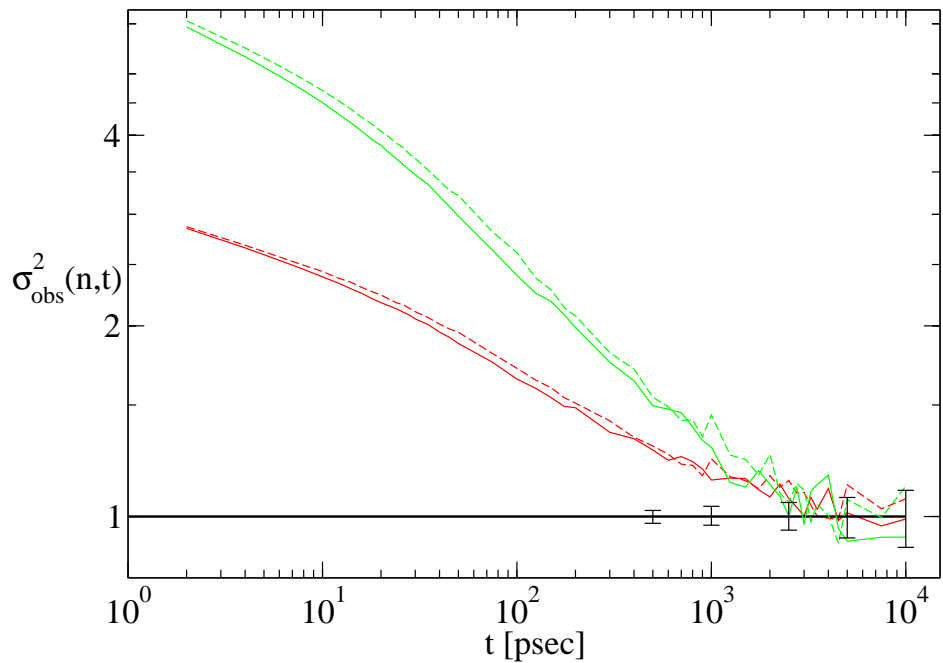}
\caption{Convergence data for two independent $1$ $\mu$sec met-enkephalin trajectories, 
distinguished by solid and dashed lines, for subsample sizes $n=4$ (red) and $n=10$ (green). 
Error bars indicate $80$ \% confidence intervals for uncorrelated subsamples of size $n=4$.}
\label{metenkfig}
\end{figure}

Despite the lack of a simple description in terms of a few, well-defined states connected by occasional 
transitions, our decorrelation analysis yields an unambiguous signal of the decorrelation time for this system. 
The data (Fig.~\ref{metenkfig}) indicate that $4$ or $5$ nsec must elapse between frames before they be considered statistically 
independent, which in turn implies that each of our $1$ $\mu$sec trajectories has an effective sample size 
of $200$ or $250$ frames. We stress that this is learned from a ``blind'' analysis, without any knowledge of 
the underlying free energy surface. 

%
%
%
%
%
%
%

\section{Discussion}
We have developed a new tool for assessing the quality of molecular simulation trajectories, quantifying  
``structural correlation'', the tendency for snapshots which are close together 
in simulation time to be similiar. The analysis first computes a structural decorrelation time, 
which answers the question, ``How much simulation time must elapse before the sampled structures display 
the statistics of an i.i.d sample.'' This in turn implies an effective sample size, ${\cal N}$,  
which is the number of frames in the trajectory that are statiscally independent, in the sense that they may 
be thought of as independent and identically distributed. 

In several model systems, for which the timescale needed to decorrelate snapshots was known in advance, we 
have shown that the decorrelation analysis is consistent with the ``built-in'' timescale. We have also 
shown that the results are not sensitive to the details of the structural histogram or to the 
subsampling scheme used to analyze the resulting timeseries. There are no adjustable parameters. Finally, we have demonstrated a 
calculation of an effective sample size for a system which cannot be approximately described in terms of 
a small number of well-defined states and a few dominant timescales. This is critically important, since 
the important states of a system are generally not known in advance.

Our method may be applied in a straightforward way to discontinuous trajectories, which consist 
of several independent pieces\cite{pande-prl01}. The analysis would be carried forward just as for a continuous trajectory. 
In this case, a few subsamples will be corrupted by the fact that they 
span the boundaries between the independent pieces. The error introduced will be minimal, provided that the correlation time is 
shorter than the length of each independent piece.

The analysis is also readily applicable to exchange-type simulations, in which configurations are swapped 
between different simulations running in parallel. For a ladder of $M$ replicas, one would perform the 
analysis on each of the $M$ \emph{continuous} trajectories that are had by following each replica as it 
wanders up and down the ladder. If the ladder is well-mixed, then all of the 
trajectories should have the same decorrelation time. And if the exchange simulation is more efficient than 
a standard simulation, then each replica will have a shorter decorrelation time 
than a ``standard'' simulation. This last observation attains considerable exigence, in light of 
the fact that exchange simulations have become the method of choice for state-of-the-art simulation.   

There is a growing sense in the modeling and simulation community of the need to standardize measures of the quality of 
simulation results\cite{archiving-struct06,quality-jctc06}. Our method, designed specifically to address the statistical 
quality of an \emph{ensemble} of structures, should be useful in this context.  

\section{Methods\label{methods}}
\subsection{Histogram Construction}
Previously, we presented an algorithm which generated a histogram based on 
clustering the trajectory with a fixed cutoff radius\cite{el-dmz-class05}, 
resulting in bins of varying probability. 
Here, we present a slightly modified procedure, which partitions the trajectory into bins of 
\emph{uniform} probability, by allowing the cutoff radius to vary. 
For a particular continuous trajectory 
of $N$ frames, the following steps are performed:

\parbox{5in}{
(i) A bin probability, or fractional occupancy $f$ is defined.\\ 
(ii) A structure $S_{1}$ is picked at random from the trajectory.\\
(iii) Compute the distance, using an appropriate metric, from $S_{1}$ to all remaining frames in the trajectory.\\ 
(iv) Order the frames according to the distance, and set aside the first $f\times N$ frames, noting 
that they have been classified with reference structure $S_1$. Note also  
the ``radius'' $r_1$ of the bin, i.e., the distance to the farthest structure classified with $S_1$.\\ 
(iv) Repeat (ii)---(iv) until every structure in the trajectory
is classified.\\
}    

\subsection{Calmodulin}
We analyzed two coarse-grained simulations of 
the N-terminal domain of calmodulin. Full details and analysis of the model have been published 
previously\cite{zuckerman-calmod}, here we briefly recount only the most relevant details. 
The model is a one bead per residue model of $72$ residues (numbers $4-75$ in pdb structure 1cfd), 
linked together as a freely jointed chain. Conformations corresponding to both the apo (pdb ID 1cfd) 
and holo (pdb ID 1cll) crystal 
structures\cite{bax-camholo-nature95,camapo-jmb92} are stabilized by 
Go interactions\cite{go-folding75}. Since both the apo and holo forms are stable, transitions 
are observed between these two states, ocurring on average about once every $5\times 10^4$ 
Monte Carlo sweeps (MCS).    

\subsection{Met-enkephalin}
We analyzed two independent $1$ $\mu$sec trajectories and a single $1$ nsec trajectory, each started from the 
PDB structure 1plw, model $1$. 
The potential energy was described by the OPLSaa potential\cite{oplsaa}, with solvation treated implicitly 
by the GB/SA method\cite{still-gbsa}. The equations of motion were integrated stochastically, using the 
discretized Langevin equation implemented in Tinker v. $4.2.2$, with a friction constant of $5$ psec$^-1$ and the 
temperature set to $298$ K\cite{tinker}. A total of $10^6$ evenly spaced configurations were stored for each 
trajectory.


\pagebreak

\pagebreak

\pagebreak

\pagebreak

\pagebreak

\pagebreak

\pagebreak

%
%
\end{document}